# A programmable electro-optic frequency comb engine for integrated parallel 3D convolutional computing

Jinze He[1,2†], Junzhe Qiang[2†], Yiying Dong[2†], Jingyi Wang[1], Tian Dong[2], Gongcheng Yue[2], Yihan Miao[2], Rongjin Zhuang[2], Mingze Lv[1], Jie Liu[1], Siyuan Yu[1], Zhongjin Lin[1], Xinlun Cai[1*], Yuanmu Yang[2*], Guanhao Wu[2*], Yang Li[1,2*]

**Affiliations:**

[1]State Key Laboratory of Optoelectronic Materials and Technologies, School of Electronics and Information Technology, Sun Yat-sen University, Guangzhou, Guangdong, China.

[2]State Key Laboratory of Precision Measurement Technology and Instruments, Department of Precision Instrument, Tsinghua University, Beijing, China.

†These authors contributed equally: Jinze He, Junzhe Qiang, Yiying Dong.

*caixlun5@mail.sysu.edu.cn; ymyang@tsinghua.edu.cn;
guanhaowu@tsinghua.edu.cn; liyang328@mail.sysu.edu.cn

**Summary**

Integrated photonic convolutional processors have emerged as a cornerstone for next-generation artificial intelligence, promising to bypass the bandwidth and energy bottlenecks of electronic hardware. However, existing wavelength-multiplexed architectures, primarily based on Kerr microcombs, are often hindered by low optical power conversion efficiency and slow weight reconfiguration speeds, which preclude their use in dynamic, high-speed computational tasks. Here, we demonstrate an electro-optic convolutional neural network (ECNN) based on a monolithic thin-film lithium niobate (TFLN) photonic processing unit. This ECNN architecture realizes multi-wavelength generation and high-speed weight mapping within a single programmable electro-optic (EO) frequency comb. By harnessing the Pockels effect and ultra-broad EO bandwidth of TFLN, our processor achieves a parallel computing speed of 1.62 trillion operations per second (TOPS) with an unprecedented weight reconstruction rate exceeding 38 GHz. We show that the system achieves near-unity optical power conversion efficiency (>99.9%), a two-order-of-magnitude improvement over state-of-the-art microcomb-based systems. Beyond static image classification, we leverage the picosecond-scale reconfigurability to demonstrate, for the first time, photonic 3D convolution for real-time video action recognition. This programmable EO comb-based framework establishes a new paradigm for monolithic photonic processors, offering a scalable and low-latency solution for autonomous systems and real-time in-sensor computing.

## Introduction

Convolutional neural networks (CNNs) are the backbone of modern artificial intelligence, enabling breakthroughs in machine vision and high-dimensional data analysis[1,2]. As electronic processors approach the physical limits of transistor scaling and energy efficiency, optical neural networks (ONNs)[3-6] have emerged as a promising alternative, offering inherent parallelism, broad bandwidth, and light-speed signal transmission. Among various architectures, wavelength-division multiplexing (WDM) systems (Fig. 1a) are particularly attractive because they exploit the frequency domain for convolution, maximizing compute density without increasing the physical footprint[7].

Despite this potential, current wavelength-multiplexed photonic convolutional processors—most notably those based on Kerr microcombss[8,9] (Fig. 1b)—face two critical limitations. First, Kerr combs typically suffer from low pump-to-comb conversion efficiency and require external spectral shaping for weight mapping, which compromises system optical power conversion efficiency (the ratio between output weighted comb lines power to input optical power) and compactness. Second, and more pivotally, weight reconfiguration in these systems often depends on slow thermo-optic phase shifters or phase-change materials. This restricted switching speed limits their application to static tasks and prevents the execution of dynamic, high-speed computational tasks, such as video processing[10,11] and in-situ training[12-14], where rapid kernel updates are essential.

Here, we introduce an electro-optic convolutional neural network (ECNN) platform that addresses these challenges by consolidating multi-wavelength generation and picosecond-scale weight mapping onto a single thin-film lithium niobate (TFLN)[15] chip (Fig. 1c). Unlike Kerr-based systems, our ECNN utilizes a programmable electro-optic (EO) frequency comb[16-18] engine, which inherently achieves near-unity power conversion efficiency (>99.9%) and exploits the giant Pockels effect for instantaneous weight updates. By mapping kernels directly onto the EO sidebands, the ECNN architecture enables a parallel computing throughput of 1.62 TOPS coupled with an unprecedented 38 GHz weight reconstruction rate.

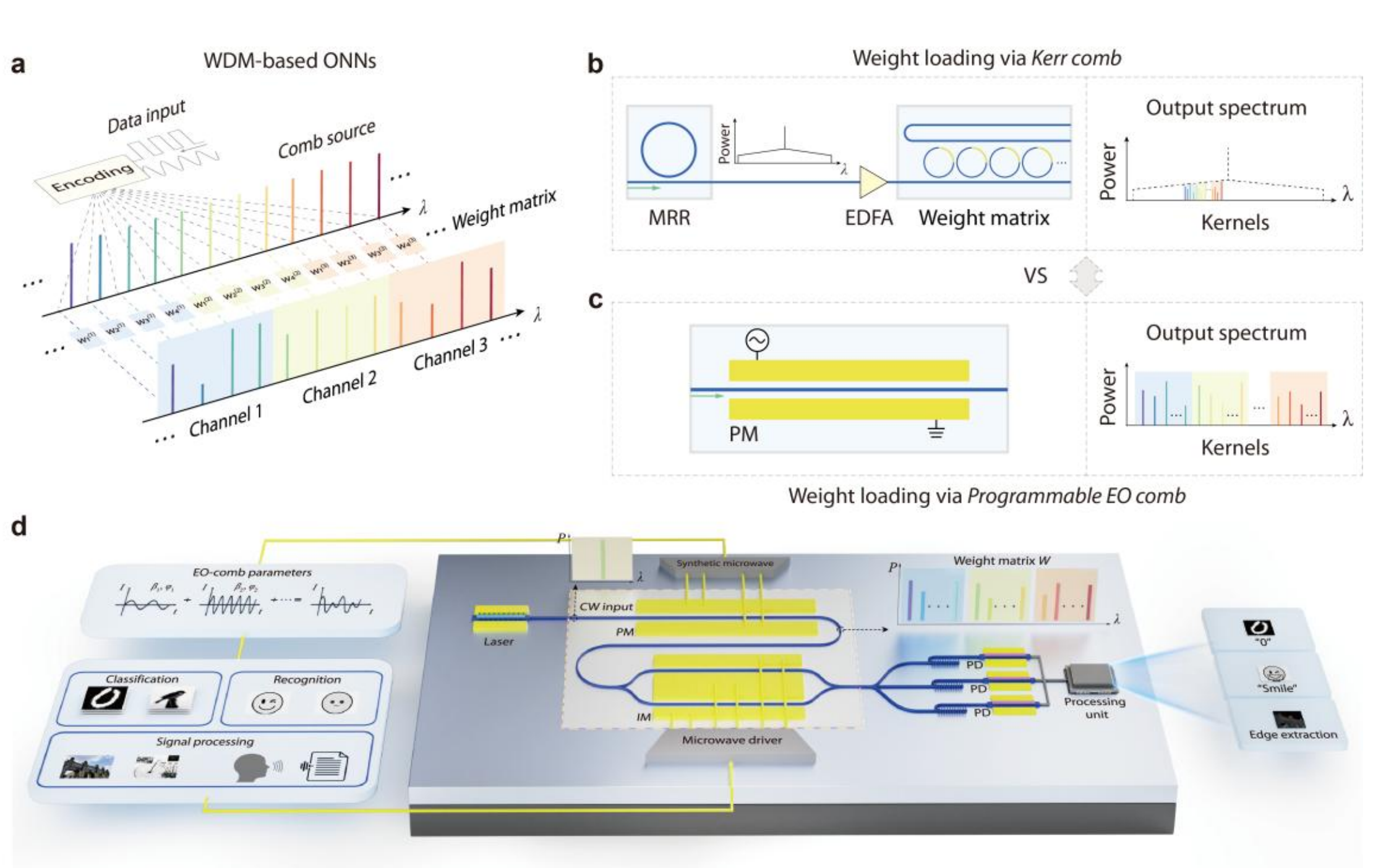

**Figure 1 | Concept of parallel convolution processing via programmable EO combs.** **a,** Schematic of wavelength multiplexing-based ONNs. **b,** Kerr comb-based weight mapping optical system. **c,** EO comb-based weight mapping optical system. **d,** Conceptual illustration of the fully integrated photonic unit used for parallel convolution processing. MRR, microring resonator; EDFA, erbium-doped fibre amplifier; PM, phase modulator; IM, intensity modulator.

To demonstrate the superiority of the ECNN paradigm, we extend photonic computing from traditional static image classification (MNIST and Fashion-MNIST) to the more demanding domain of 3D convolutional video action recognition[19] (KTH dataset[20]). We show that the picosecond-scale reconfiguration latency allows the processor to extract spatio-temporal features from continuous video frames in real-time, a feat unattainable by prior photonic architectures. This monolithic ECNN framework not only optimizes the energy-speed-footprint trade-off but also provides a scalable route toward high-performance, in-sensor photonic intelligence.

## Integrated 3D parallel convolution processing unit

The 3D integrated parallel convolution processing unit consists of a phase modulator and an intensity modulator monolithically integrated on a TFLN chip (Fig. 1d). Under the optical pump from a tunable continuous-wave laser, the phase modulator is driven by the fundamental microwave tone $\omega_{\mathrm{m}}$ and its multiple higher-order harmonics. Each harmonic is independently tunable, corresponding to different phase modulation indices $\beta$ and initial phases $\phi$. For distinct computational tasks, we optimize or train corresponding parameters, which are subsequently encoded into the phase modulator to generate EO combs serving as weight matrices. The complex amplitude of the *p*th-order comb line away from the centre frequency is written as equation (1)

$$E(p)=E_0\sum_{p_1=-\infty}^{\infty}\sum_{p_2=-\infty}^{\infty}\cdots\sum_{p_q=-\infty}^{\infty}J_{p_1}(\beta_1)J_{p_2}(\beta_2)\cdots J_{p_q}(\beta_q)e^{i(p_1+2p_2+\cdots+qp_q)\omega_{\mathrm{m}}t+i(p_1\phi_1+p_2\phi_2+\cdots+p_q\phi_q)}\delta(p-p_1-2p_2-\cdots-qp_q), \quad (1)$$

where $E_0$ is the amplitude of the input optical field and $q$ is the number of harmonics involved in the modulation. $J_p$ and $\delta$ refer to the *p*th-order Bessel function of the first kind and the Dirac delta function, respectively (see Supplementary information 1 for details). Distinct from EO comb generated by single-tone modulation, where the coupling is localized between adjacent comb lines and the intensity distribution is fixed to be in the form of the Bessel function, the presence of higher-order harmonics introduces long-range coupling between non-adjacent comb lines. And, the initial phases of different microwave harmonics also influence the constructive and destructive interferences between different coupling states, which are absent in the case of single-tone modulation[21,22]. By programming the phase modulation index and initial phase of each harmonic, we can tailor the optical power of individual EO comb lines, which in turn encodes the weights of multiple parallel convolutional weight matrices **W**. The input data vector **X** is then encoded into temporal microwave signals and mapped into each kernel comb line via the intensity modulator. The optical signal output from the TFLN chip passes through a delay line to ensure the alignment of the convolutional unit with the corresponding coded data vector element. Subsequent photodetector (PD) implements the sum of local data matched to each convolutional weight.

Figures 2a-d show the photograph and the microscope image of the packaged parallel convolution processing unit (see method and Supplementary information 2 for details). The device employed two cascaded folded phase modulators with capacitance-loaded traveling-wave electrodes for EO comb generation. Compared to the conventional single-pass phase modulator, a triple-pass configuration increased the modulation length via waveguide crossings, leading to a lower driving voltage while maintaining a compact footprint. We characterized the half-wave voltage of the cascaded folded phase modulator without a push-pull configuration by analysing the EO comb spectra (Fig. 2e). The measured half-wave voltages were as low as 2.61 V at 19.08 GHz and 6.02 V at 38.16 GHz. Then, a folded intensity modulator in a travelling-wave manner was used to receive the input data vector. The double-pass architecture minimized the half-wave voltage while ensuring a large EO bandwidth. As shown in Figs. 2f and g, with a 1.75-cm modulation length, the intensity modulator exhibited an ultra-low half-wave voltage of 1.27 V (Fig. 2f) and a 3-dB EO bandwidth exceeding 35 GHz (Fig. 2g). At a 30-gigabaud (GBaud) data rate with a root-mean-square voltage of 140 mV, the intensity modulator showed an ultra-low electrical energy consumption of 13 fJ/bit.

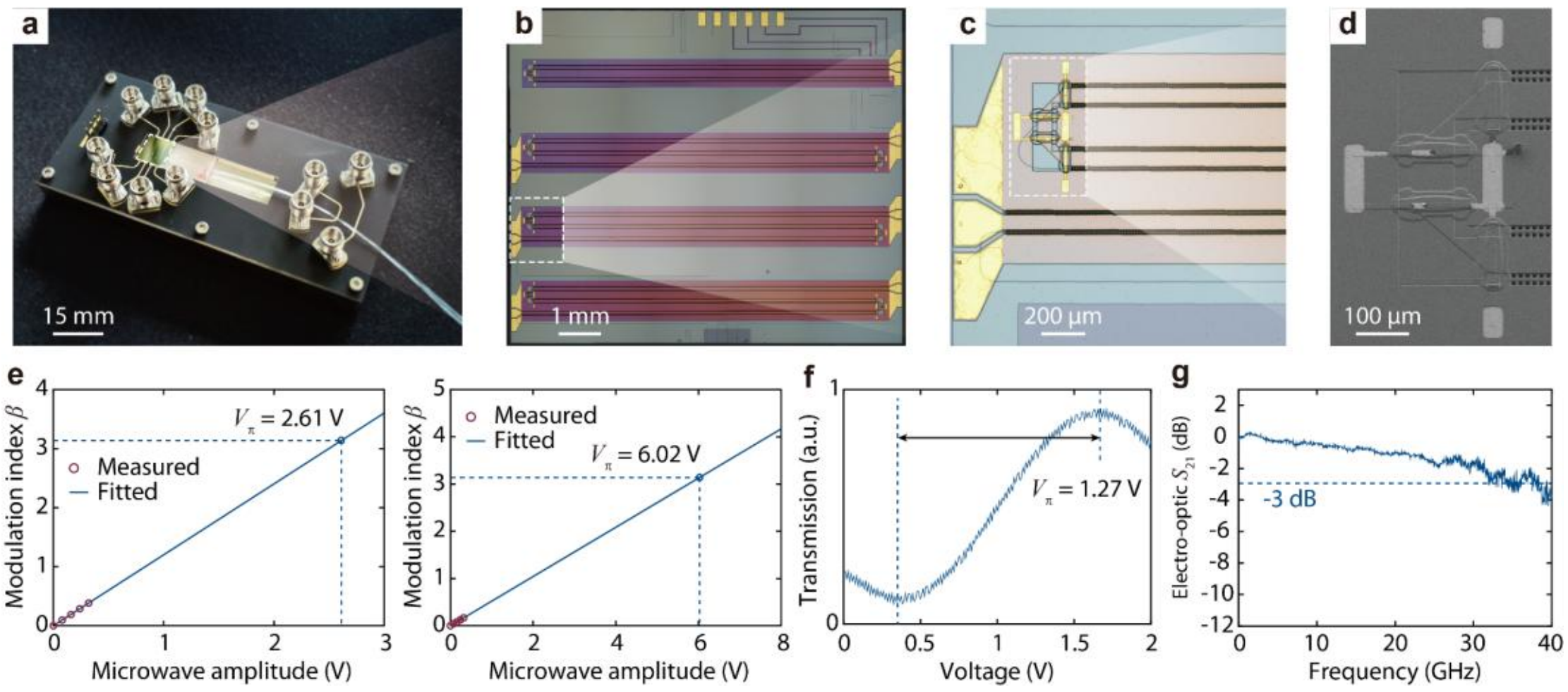

**Figure 2 | Device images and performances. a,** Photograph of the packaged integrated photonic unit. **b-d,** Optical microscope images (**b, c**) and scanning electron microscope image (**d**) of the integrated photonic unit. The zoomed-in images in **c** and **d** illustrate the U-turn bend, capacitance-loaded traveling-wave electrode, and RF wire bonding of

the folded modulator. **e.** Phase modulation index fitting of the phase modulator, indicating half-wave voltages of 2.61 V at 19.08 GHz and 6.02 V at 38.16 GHz. **f,** Normalized transmission of the intensity modulator as a function of the applied voltage, showing an ultra-low half-wave voltage of 1.27 V. **g.** Measured EO response of the intensity modulator. The 3-dB bandwidth is broader than 35 GHz.

**Programmable EO comb for convolution processing**

To demonstrate the programmability and precision of EO combs, we employed optimization algorithms to tailor convolutional kernels for implementing image processing. The spectral profile of programmable EO combs was determined by phase modulation indices $\beta$ and initial phases $\phi$. Figure 3a illustrates the particle swarm algorithm loop for optimizing these two parameters, considering only the fundamental tone and the second harmonic. First, the power of each comb line is calculated using the Fast Fourier Transform (FFT) method. Then, the fitness is evaluated by the root mean square error (RMSE) between the target comb and the calculated FFT spectrum. The optimization loop terminates and outputs the optimized parameters once the fitness threshold is reached. Otherwise, new position (including personal $P_{best}$ and global best parameters $G_{best}$) and velocity (parameter change rates) are updated for the next iteration. The output phase modulation indices $\beta$ are converted to microwave drive signal's peak amplitudes $V_p$ via $\beta = \pi V_p / V_\pi$, where $V_\pi$ is the half-wave voltage of the phase modulator. Finally, the fundamental tone and second harmonic carrying specific initial phases are synthesized by an arbitrary waveform generator (AWG) and amplified before being applied to the phase modulator.

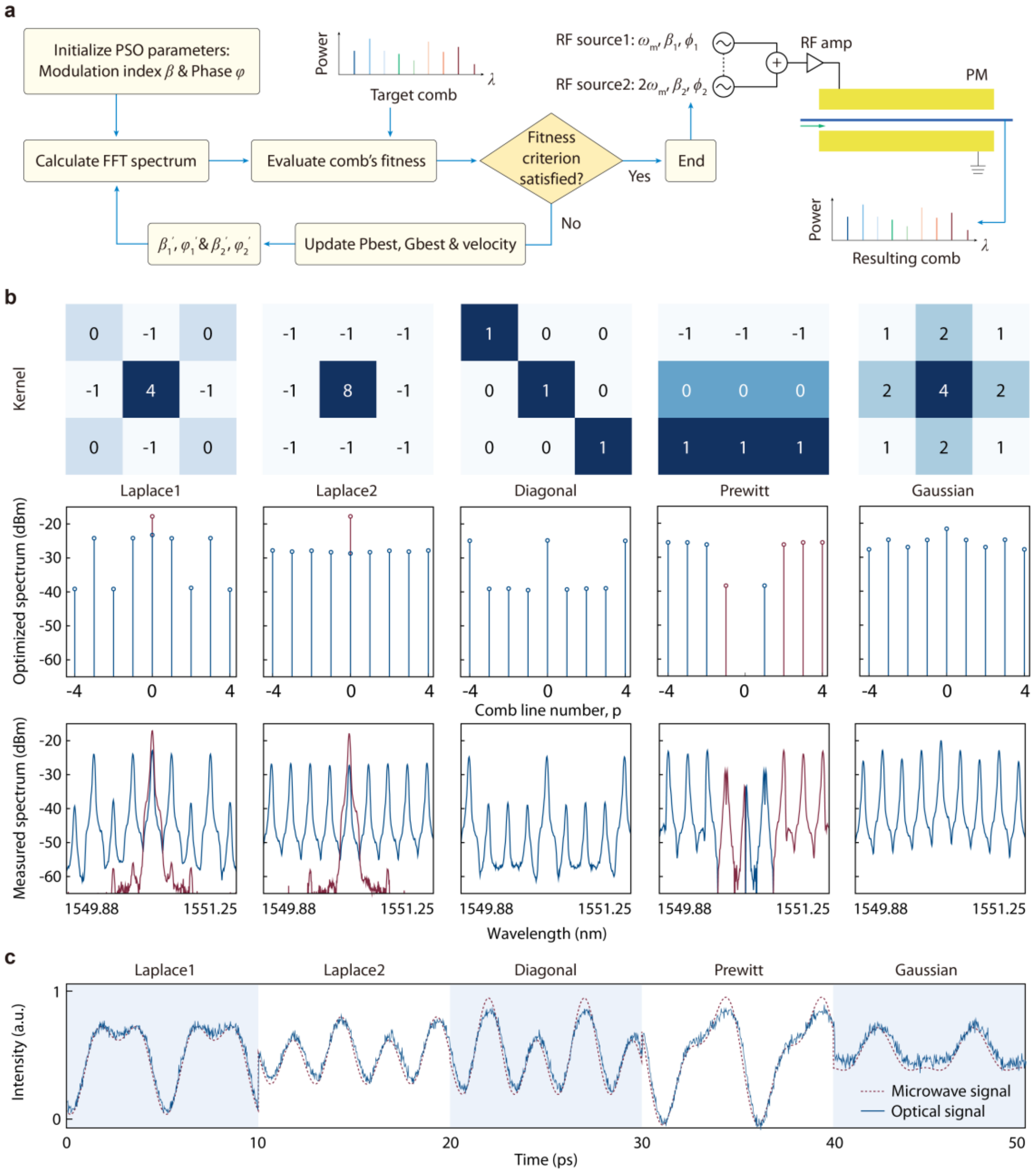

**Figure 3 | Optimization and generation of programmable EO combs. a,** Flowchart of the particle swarm algorithm used to optimize phase modulation indices and initial phases of microwaves and in turn to match the target comb spectral profile. **b,** The optimized optical spectra (middle row) and the corresponding measured optical spectra (bottom row) of five weight matrices (top row). **c,** The measured optical temporal waveform driven by the designed microwave with a modulation rate of 30 GBaud. Amp, amplifier.

Five standard 3×3 image-processing convolutional kernels shown in the top row of Fig. 3b were selected to validate the feasibility of programmable EO comb. The second row in Fig. 3b shows the optimized spectra for each target kernel with RMSEs of [0.0062, 0.0055, 0.0347, 0.0819, 0.0151] between the tailored EO combs and target kernel weights. To experimentally verify these results, we drove the phase modulator with synthetic microwaves consisting of a 19.42-GHz sinusoid and a 38.84-GHz sinusoid, yielding the measured spectra in the bottom row of Fig. 3b. The experimental RMSE values between the measured EO comb spectra and target kernel weights are [0.0054, 0.0053, 0.0336, 0.0655, 0.0269], which match well with the optimized spectra and give a weight precision up to 6.5 bit (see Supplementary information 3 for details). The precise alignment of EO combs indicates a modulation speed and thus a kernel switching speed exceeding 38 GHz (repetition rate). To further demonstrate the ultra-high switching speed of the convolutional kernel, we applied the temporal synthetic microwave waveforms (encoding all five target kernels) to the modulator at a modulation rate of 30 GBaud (Fig. 3c), yielding distortion-free time-domain signals and in turn confirming the ultra-high weight reconstruction speed.

To demonstrate the convolutional computing capability of the programmable EO comb, we implemented image edge detection utilizing a 3×3 Laplace kernel (Fig. 4a). First, the Laplace kernel matrix **W** can be decomposed into a linear subtraction of positive ($\mathbf{W_p}$) and negative ($\mathbf{W_n}$) channels via the distributive property of convolution operations (see Supplementary information 4 for details). The corresponding programmable EO combs for $\mathbf{W_p}$ and $\mathbf{W_n}$ were sequentially generated using the phase modulator. Next, the data matrix of the input image was partitioned into convolution-sized units and flattened into a one-dimensional data vector for optical processing. In our experiment, the input 512×512 8-bit grayscale image was transformed into a 1×783,360 input vector **X** and then loaded onto the intensity modulator through the AWG operating at a modulation rate $f$ of 30 GBaud. Each kernel comb line consequently carried the complete information required for the convolution operation. Then, the EO comb propagated through a dispersion-compensating fibre with a dispersion of −213.8 ps/nm, which can introduce a time delay $\Delta t$ of 33.3 ps between adjacent kernel comb lines. The time delay $\Delta t$ was determined by $\Delta t = \frac{\mathrm{c}}{f_{\mathrm{R}} \times \lambda^2 \times f}$, where c is the speed of light, $f_R = 19.08$ GHz is the repetition rate of the EO comb, and $\lambda$ = 1550.17 nm is the pump wavelength. The resulting temporal alignment ensured each kernel weight interacted with its corresponding encoded data element. Finally, the kernel comb lines were extracted by an optical fibre filter and detected by a high-speed PD, which performed the weighted sum of the data vector. To match the sliding-window operation of the convolution, the sampling rate of PD was set to $1/N$ of the encoding rate ($N = 3$ is the kernel size).

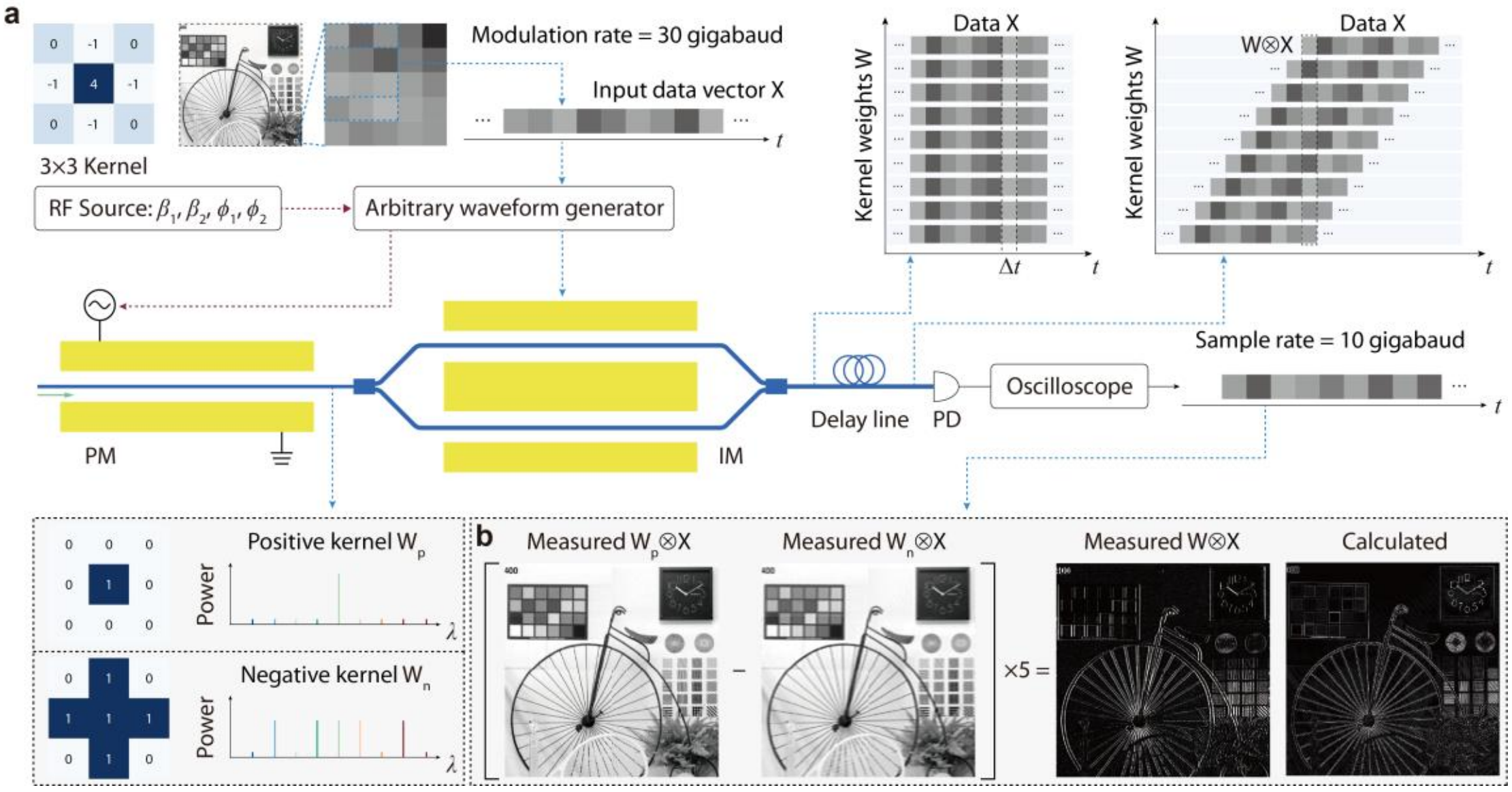

**Figure 4 | Image edge detection. a,** Operation principle of convolution for image edge detection with two 3×3 kernels (decomposition of Laplace operator). **b,** Images recovered after convolution.

In the experiment, we first obtained the convolution results between the weight matrices ($\mathbf{W_p}$, $\mathbf{W_n}$) and the input vector **X**. Leveraging the distributive property of convolution, the normalized intensity difference between these two images — amplified by a factor of five — yielded the convolution between the Laplace kernel **W** and the data vector **X**. The measured results exhibited excellent agreement with theoretical calculations, confirming the feasibility and accuracy of the programmable EO comb-based convolution operations (Fig. 4b).

**ECNN training framework for image classification**

The programmability of the EO comb provides great flexibility, allowing us to perform not only image processing guided by specific kernels but also computational task-oriented image classification. We designed an ECNN training framework (Fig. 5a) consisting of an optical convolutional layer followed by two fully connected layers. The optical convolutional layer employed three parallel 3×3 kernels implemented by 27 comb lines of the programmable EO comb. The spectral profile of the EO comb was determined by the phase modulation indices and initial phases of the fundamental tone and second harmonic, allowing us to replace kernel weights with four parameters $(\beta_1', \phi_1', \beta_2', \phi_2')$ as variables (Fig. 5a). For different training tasks, we input the corresponding image with a 28×28 pixel matrix. After convolution with the three parallel 3×3 kernels, we generated three 26×26 feature maps, which were then flattened into a 1×2028 vector. Such a vector subsequently passed through two fully connected layers (2028×256 and 256×10) thus accomplishing classification. Because the optical convolutional layer, as a function of the phase modulation indices/initial phases, is differentiable, parameter training in this framework followed the standard backpropagation algorithm. The trained phase modulation indices and initial phases can be directly loaded onto the phase modulator to generate the required programmable EO comb for the classification task.

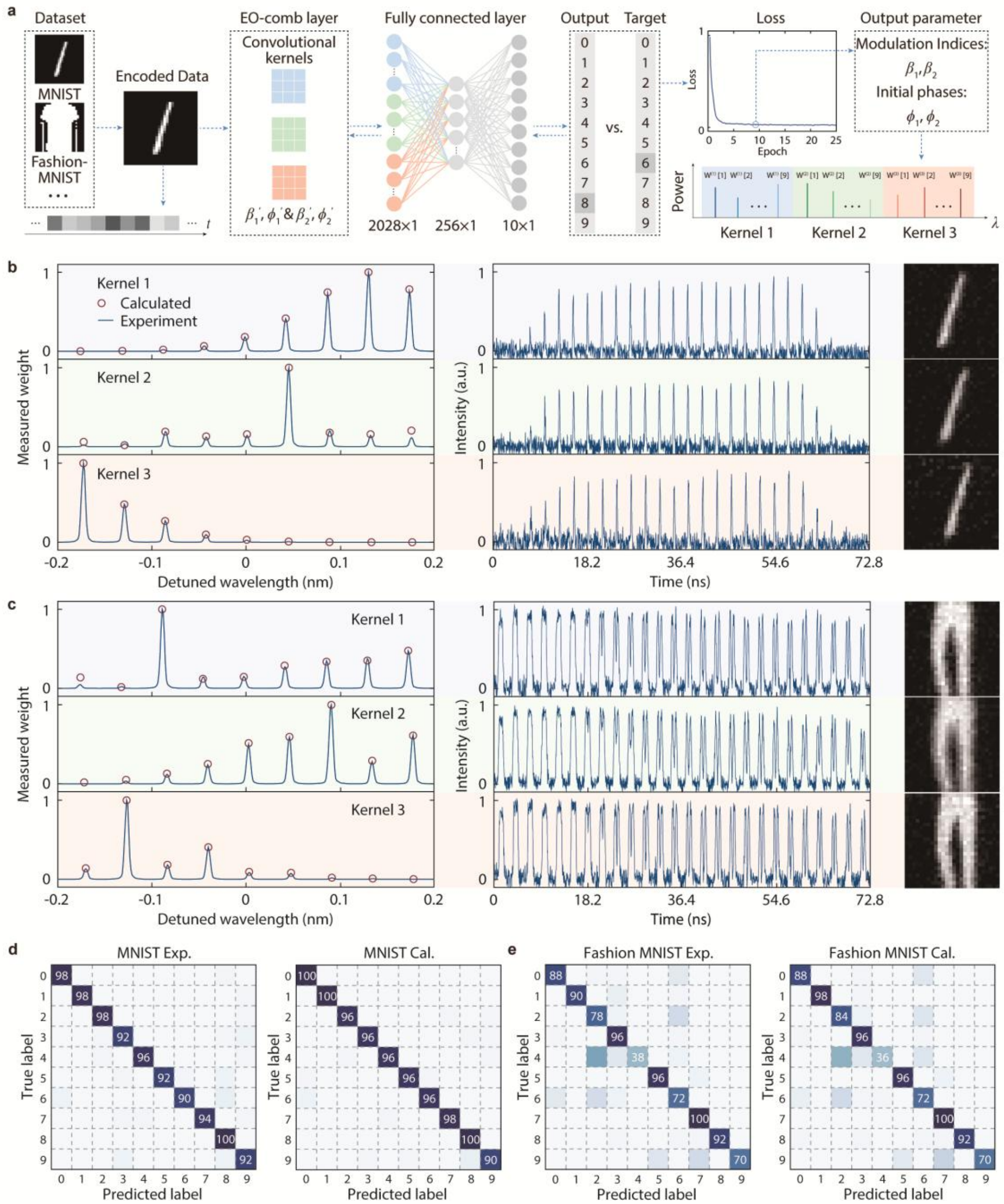

**Figure 5 | Image recognition processed by parallel convolution. a,** Flowchart of the ECNN training framework to optimize the comb spectral profile matching the convolutional layer of image recognition CNNs. **b-c,** The first column shows calculated and measured normalized spectra of three parallel kernels used for classification of the **b** MNIST dataset, and **c** Fashion-MNIST dataset. The second and third columns are the corresponding temporal waveforms and recovered feature maps. **d-e,** The experimental and calculated confusion matrices for classification of the **d** MNIST dataset, and **e** Fashion-MNIST dataset.

We performed ECNN training for the MNIST dataset to obtain the phase modulation indices [4.8877, 1.8567] and phases [4.4844, 4.7707], which were then used to generate the EO comb (see Supplementary information 5 for details). The central 27 comb lines were divided into three parallel convolutional kernels, as shown in the first column of Fig. 5b (see Supplementary information 6 for details). The measured weight matrices exhibited a low average RMSE of 0.021 compared with the theoretical weights. We experimentally evaluated the classification accuracy using 500 test images. The second and third columns of Fig. 5b present the temporal waveforms and recovered feature maps for the three kernels, respectively. For a specific category, analysis of 50 images (totalling 33,800 sampling points) yielded a standard deviation of 0.051 between theoretical and experimental results (see Supplementary information 6 for details). This discrepancy was primarily attributed to the low signal-to-noise ratio caused by weak output optical signals. The confusion matrix (Fig. 5d) reveals experimental and theoretical accuracies for the MNIST dataset of 95.0% and 96.8%, respectively.

We further extended our approach to the Fashion-MNIST dataset. As shown in the first column of Fig. 5c, the measured kernel weights achieved an average RMSE of 0.02 compared with the theoretical weights. The corresponding confusion matrix (Fig. 5e) shows an accuracy of 82.0% for the experiment versus 83.2% for the calculation. The excellent agreement between theoretical and experimental accuracies of both MNIST and Fashion-MNIST datasets-based classification tasks demonstrates that our ECNN-trained EO comb can be universally adapted to diverse CNNs. Furthermore, the optical power conversion efficiency of our programmable EO comb-based kernel exceeded 99.9%, representing a two-order-of-magnitude improvement over the conventional Kerr comb-based kernels (see Supplementary information 7 for details). With three parallel kernels and a data rate of 30 GBaud, the computing speed of the photonic core reached 1.62 TOPS ($3\times2\times9\times30\times10^9$).

**ECNN training framework for video recognition**

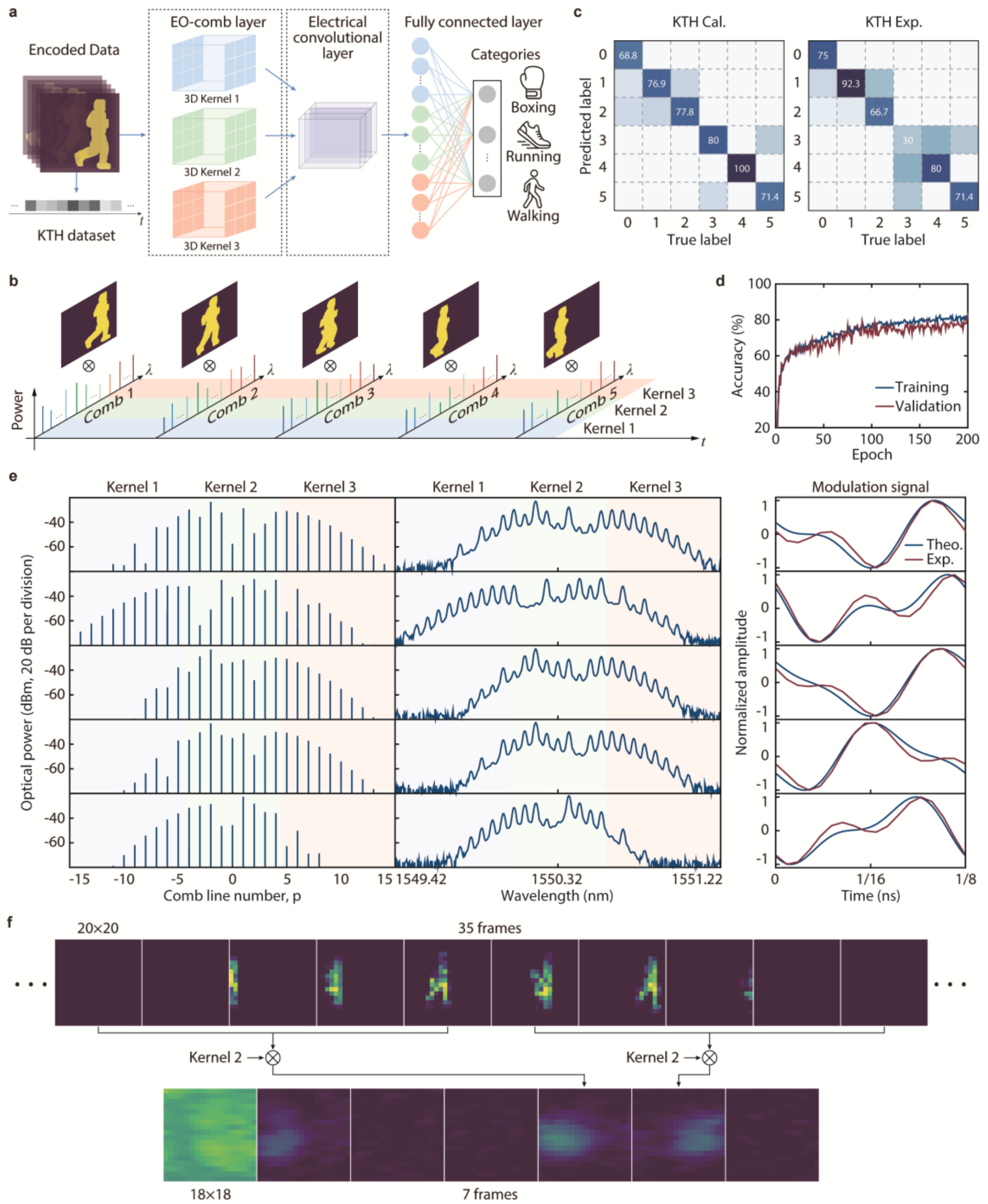

**Figure 6 | Video recognition processed by 3D convolution. a,** Flowchart of the 3D ECNN training framework to optimize the comb spectral profile matching the convolutional layer of video recognition CNNs. **b,** Operation principle of parallel 3D convolution for video recognition with five distinct programmable EO combs. **c,** The calculated and measured confusion matrices for classification of the KTH dataset. **d,** Training accuracy and validation accuracy of the 3D ECNN. **e,** The left and middle columns show calculated and measured spectra of three parallel 3D kernels used for classification of the KTH dataset. The right column shows the corresponding theoretical

and calibrated temporal waveforms loaded to the AWG. **f,** Recovered feature maps obtained by applying the convolution kernel2 to a 14-frame running motion video sequence.

In addition to image classification tasks implemented via fixed-programmable EO comb convolutional layers, the ultra-fast switching capability of programmable EO combs allows their extension to higher-dimensional CNNs for more complex tasks, such as video-based action recognition. Figure 6a illustrates a 3D ECNN designed for action recognition on the KTH dataset. The video sequence is processed in temporal blocks of five frames, each block corresponding to five distinct programmable EO combs (Fig. 6b). Similar to the image classification setup, each programmable EO comb utilizes its central 27 comb teeth in the convolution, thereby forming three parallel convolutional kernels of size 5×3×3. After undergoing convolution and summation operations, each video is transformed into an 18×18×7 tri-pixel matrix. To enhance the performance of the 3D ECNN, the optical convolutional layer is followed by an electronic convolutional layer (with 16 kernels of size 5×3×3) and two fully-connected layers (16×32 and 32×6). A 2×2 max pooling layer is appended after each convolutional layer, and dropout is applied after pooling during training to mitigate overfitting. As shown in Figs. 6c-d, the 3D ECNN achieves a theoretical validation accuracy of 76.7 % (see Supplementary information 10 for details). The theoretical spectra of the five programmable EO combs are presented in the left column of Fig. 6e (see Supplementary information 10 for details).

In the experiment, to match the 120-GHz sampling rate of the AWG, we selected a repetition rate of 8 GHz for the programmable EO comb (encoded by combining an 8-GHz fundamental tone and a 16-GHz harmonic). Consequently, each comb period could be described with 15 sampling points. Considering the amplitude and phase errors between the fundamental and harmonic microwaves introduced by the microwave power amplifier and traveling-wave electrodes (measured amplitude error: $V_{16GHz}/V_{8GHz} = 1.85$, phase error: $\phi_{16GHz}$-$\phi_{8GHz} = -0.2$ rad), calibration was performed on the time-domain coding of each programmable EO comb. The corresponding normalized theoretical and experimental encoded waveforms are shown in the right column of Fig. 6e. Furthermore, we employed a 16-GHz data encoding rate, ensuring each video frame corresponds to 600 (20×20×3×8/16, where 3 is the kernel size) integer modulation cycles of the programmable EO comb. This guarantees each frame is convolved with an independent and stable programmable EO comb. Ultimately, the five programmable EO combs generated in the experiment are shown in the middle column of Fig. 6e, with corresponding RMSE of [0.00127, 0.0137, 0.0069, 0.0087, 0.0432] and a power conversion efficiency exceeding 99.9%.

Owing to the limited bandwidth of the optical filter and signal-to-noise ratio, only the convolution data corresponding to the second kernel (kernel 2) were collected in the experiment (see Supplementary information 11 for details). The convolution results of kernel 2 for a 14-frame video sequence of a running action are presented in Fig. 6f. Feeding the experimental results into the back-end electronic network illustrated in Fig. 6a yielded an action-recognition accuracy of 70% on the KTH dataset. With the 16-GHz encoding rate in the optical convolution stage, the processing speed for a frame of 20×20 pixels reaches 13.3 MHz. Even considering a typical 4K video (8,294,400 pixels per frame) requiring a refresh rate of 240 Hz, our processing speed as high as 643 Hz can still meet this requirement.

## Discussion

The TFLN-based programmable EO comb processor presented here addresses the critical trade-offs between efficiency, speed, and integration in optical computing. By comparing our device with state-of-the-art integrated photonic computing frameworks (Table 1), several distinct advantages emerge. First, the intrinsic second-order nonlinearity of the EO comb permitted a near-unity optical power conversion efficiency (>99.9%), which is two orders of magnitude higher than that of the Kerr comb. The EO comb-based parallel convolution operation thus fully leveraged the broad optical bandwidth of ONNs, resolving the challenge of stringent power requirements for on-chip light sources and system robustness.

**Table 1:** Comparison of state-of-the-art integrated photonic processing units

| Type | Platform | Source | Conversion efficiency | Kernel number | Matrix dimension | Kernel switching speed (GHz) | Modulation baud rate (Gbaud) | Compute speed (TOPS) |
|---|---|---|---|---|---|---|---|---|
| This work | TFLN | EO comb | >99.9% | 3 | 3×9 | >38 | 30 | 1.620 |
| WDM[23] | SiN + EOM[a] + PCM | Kerr; BS | <0.98%[b] | 4 | 4×9 | $\sim 1\times10^{-5}$[c] | 2 | 0.576 |
| WDM[24] | AlGaAs + SOI | Kerr; DS | <0.22%[b] | 1 | 1×4 | $<3\times10^{-4}$[d] | 17 | 0.136 |
| MMI[25] | EOM[a] + SiN | 4×CW | / | 4 | 1×4 | $\sim 2\times10^{-4}$[e] | 16.6 | 0.531 |
| MZI[26] | EOM[a] + SOI | 1×CW | / | 4 | 2×2 | $<3\times10^{-4}$[d] | 40 | 1.28 |
| MZI[27] | TFLN | 1×CW | / | / | 4×4 | >20 | 20 | 0.640 |

[a]Commercial EO intensity modulator. [b]Extracted from the spectra in ref. [23,24] (see the Supplementary information 7 for details). [c]Estimated via ref. [28]. [d]We referred to the state-of-the-art silicon thermo-optic phase shifters in ref. [29]. [e]Referred to the device in ref.[30]. WDM, wavelength division multiplexing; EOM, electro-optic modulator; PCM, phase-change material; BS, bright soliton; DS, dark soliton; CW, continuous wave.

Secondly, the compact TFLN processing unit enabled scalable convolution processing with high computing speed. The traveling-wave electrode configuration of the intensity and phase modulators ensured both microwave-optical group velocity matching and high modulation efficiency, enabling ultra-broad EO bandwidth. This allowed us to generate an EO comb featuring over 30 comb lines with a repetition rate of 19.08 GHz, supporting parallel convolution operations with 1.62-TOPS computing speed at a 30-GBaud data rate. By fully leveraging the over 110-GHz bandwidth[31] of the state-of-the-art TFLN intensity modulator as well as an EO comb featuring over 60 comb lines and a higher repetition rate generated with a driving microwave with higher frequency and amplitude, our system could be scaled to six parallel 3×3 kernels at 100-GBaud encoding speed, leading to a computing speed exceeding 10.8 TOPS. By integrating our photonic core with a high point-rate LiDAR device[32] on a single TFLN chip, we could achieve a monolithic sensing and computing unit with an unprecedented throughput.

Thirdly, we proposed an ECNN training framework that can dynamically generate task-specific phase modulation indices and phases. We experimentally demonstrated classification accuracies competitive with digital computing benchmarks on both MNIST and Fashion-MNIST datasets. Simulation results further indicated the programmable EO comb's capability to handle more complex datasets (e.g., CIFAR, see Supplementary information 8 for details). Besides, the photonic unit achieved a kernel switching speed over 38 GHz, enabled by the large bandwidth of the modulator, with the theoretical limit approaching the EO bandwidth limit of TFLN modulators. This picosecond-scale reconfigurability is not merely a quantitative improvement; it is an enabler for higher-dimensional computing. As demonstrated by the KTH dataset experiments, our system can dynamically update weights to process temporal blocks of video frames in real-time, a task that would be infeasible with slower modulation mechanisms.

Experimental power consumption was mainly caused by bench-top instruments, including pump laser, EDFA, microwave source, and electrical digital computer, totalling about 25.59 W (~25 W from bench-top instruments, see Supplementary information 9 for details). Through monolithic integration of the heterogeneous laser[33,34], on-chip microwave driver[35], silicon photodetector[36], and digital and analogue circuit block, combined with low insertion loss (current insertion loss around 30 dB), the system could achieve a total power consumption as low as 840.93 mW, yielding energy efficiencies of 1.93 TOPS/W at a computing speed of 1.62 TOPS and 1.97 TOPS/W at a computing speed of 10.8 TOPS (see Supplementary information 9 for details). Our photonic core occupied a footprint of nearly 19.44 mm², yielding a compute density of 0.083 TOPS/mm² (see Supplementary information 9 for details).

Furthermore, the dispersion-compensating fibre can be replaced by a TFLN chirped Bragg grating waveguide[37] (see Supplementary information 12 for details). A fully integrated architecture could have a footprint of around 19.83 mm², delivering a potential compute density of 0.54 TOPS/mm².

The degradation of the optical signal-to-noise ratio in our experiments originated from two main sources. First, the weak output optical signal, which could be enhanced by reducing the insertion loss of the photonic core. Second, the phase noise which might be induced by either the microwave amplifier or the beat signal of the EO combs. This phase noise might be mitigated by employing a low-noise microwave amplifier and increasing the repetition rate of the EO comb.

## Conclusion

In summary, we demonstrated a TFLN photonic processing unit harnessing programmable EO combs as both a multi-wavelength source and convolutional weights. The novel EO comb source enabled convolution processing with superior compactness, throughput, and scalability. Benefiting from the high nonlinear conversion efficiency and broad EO bandwidth of integrated TFLN EO modulators, a CMOS-compatible optical pump is sufficient to generate weight matrices with an ultra-high reconstruction speed. This picosecond-scale reconfigurability allows the system to transcend static image processing, enabling dynamic, high-speed computational tasks such as 3D convolution for video action recognition and online learning. Our approach introduced the concept of programmable EO combs to the photonic processing unit, leading to compact and low-latency wavelength multiplexing-based ONNs with various applications from real-time machine vision to natural language processing.

## Methods

### Device fabrication and characterization

The photonics processing unit was fabricated on 675 μm x-cut TFLN wafers (NANOLN) with a lithium-niobate layer thickness of 360 nm and a buried silicon-dioxide layer thickness of 9 μm. Patterning was performed via deep-ultraviolet (DUV) lithography (180 nm resolution), followed by inductively coupled plasma (ICP) etching to form waveguide structures. To achieve efficient EO modulation and low-loss fibre coupling, we employed a two-step etching process. First, rib waveguides were partially etched (180 nm depth) to optimize modulation efficiency. Second, a full etch defined narrow strip waveguides in the remaining 180 nm of lithium niobate for efficient edge coupling. A 1.2-μm-thick $SiO_2$ cladding was then deposited via plasma-enhanced chemical vapor deposition (PECVD), serving as a low-loss interlayer to prevent metal-induced absorption in regions where waveguides and electrodes overlap.

For the electro-optic modulators, we defined 23-μm-wide windows in the cladding using DUV lithography and etched 200 nm of $SiO_2$ to expose the lithium-niobate surface. This design enabled strong EO interaction via a narrow T-rail gap. Next, a 100-nm-thick NiCr layer was deposited and patterned to function as a thermo-optic phase shifter and an on-chip resistor for the MZI. Subsequently, 900-nm-thick gold RF electrodes were formed via a lift-off process. A protective $SiO_2$ layer was then deposited over the wafer. To minimize microwave losses and ensure velocity matching between optical and RF signals, we performed a deep etch through the $SiO_2$ cladding, lithium-niobate film, and buried oxide, followed by a dry etch to remove a 25-μm-thick portion of the silicon substrate beneath the RF electrodes. More fabrication details can be found in our previous work[38].

The half-wave voltage of the phase modulator was obtained by fitting modulation indices of EO comb spectra, where the spectra were measured by an optical spectrum analyser (Yokogawa AQ6370D). The 3-dB EO bandwidth of the intensity modulator was measured by a vector network analyser (Keysight N5227B), and the half-wave voltage of the intensity modulator was measured by applying a 100 kHz triangular voltage. The insertion loss of the packaged photonic processing unit is about 30 dB.

**Details of the image edge detection**

The synthetic microwave signal driving the phase modulators was generated by two channels of an AWG (Keysight M8199A). The outputs were combined via a microwave power combiner and then amplified by a microwave power amplifier (Talent Microwave TLPA18G40G-45-40-HS). Next, the amplified signal passed through the second beam splitter and was loaded onto the two cascaded phase modulators. For data encoding, the 512×512-pixel image was partitioned into eight segments due to AWG waveform memory (512 KSa per channel) constraints. Each segment was sequentially encoded and output via the third AWG channel to drive the intensity modulator. The input light was provided by a tunable continuous-wave laser (Keysight N7776C), which was polarization-controlled before coupling into the photonic processing unit. The output light was transmitted through a dispersion-compensating fibre with a dispersion of −213.8 ps/nm and then filtered by a fibre Bragg grating with a 3-dB bandwidth of 1.285 nm (centred at 1550.127 nm). The filtered signal was split by a 10:90 coupler, where 10% was monitored by the optical spectrum analyser in real-time and the rest 90% power was detected by a high-speed PD (Finisar XPDV3120R). The RF signal was finally captured by a high-definition serial data analyser (Teledyne LeCroy SDA 8330D). To mitigate phase noise induced by EO combs, the analyser's sampling bandwidth was set to 10 GHz, with post-processing performed at an analogue bandwidth of 3 GHz.

**Parallel convolution operation of ECNN**

The input ten-class dataset was divided into ten subsets, which were sequentially encoded and output by the AWG and loaded onto the intensity modulator. To demonstrate parallel convolution operations using the programmable EO comb, we divided the 27-line EO comb into three groups by dynamically tuning the pump laser's centre wavelength (with a step size matching the channel bandwidth). Each comb group was then filtered by the fibre Bragg grating. The image computing speed of the processing unit was 1.62 TOPS/3 = 0.54 TOPS because of the matrix flattening overhead of 3 (kernel size of 3×3 kernels). The encoding duration of each image was 28×28×3/30 GHz = 78.4 ns, corresponding to an optical image processing rate of 1/78.4 ns = 12.8 million dataset images per second.

**Convolution operation of 3D ECNN**

Before training the 3D ECNN via the electrical terminal, the KTH dataset was preprocessed. The KTH video dataset was categorized into six action classes and moved into corresponding subfolders. From these, videos were randomly selected to form training, validation, and test sets, comprising 300, 122, and 100 videos, respectively. The foreground of each video frame was extracted using the MOG2 background subtraction algorithm, followed by a morphological opening operation to remove noise. Subsequently, the video frames were resized to 20×20 pixels, and a fixed number of frames (35 frames) were sampled at uniform intervals. The background-subtracted videos were normalized and horizontally flipped to enhance generalization. Finally, the preprocessed training, validation, and test sets exhibited shapes of (600, 35, 20, 20, 1), (244, 35, 20, 20, 1), and (200, 35, 20, 20, 1), respectively.

During training, the modulation coefficients and phase parameters of the optical comb convolution layer were constrained to the ranges of [1, 4.5], [1, 4], and [0, 2π], respectively. Due to the distinct parameter characteristics between the optical and electrical network layers, a custom optimizer was employed, with learning rates set to 0.005 and 0.0015 for the optical and electrical layers, respectively. Training was conducted with a batch size of 15 over 200 epochs. In contrast to the dispersion value of −213.8 ps/nm used in image classification tasks, the dispersion delay was set to −975.9 ps/nm in the video classification task to match the 8 GHz repetition rate of the EO comb. The output data after convolution were reshaped into a format of (35, 18, 18, 1), then summed and transformed into (7, 20, 20, 1), before being fed into the pre-trained electrical backend network to complete the final classification.

**Code availability**

The Python codes used in this paper are available at https://github.com/hjz981125-

png/ECNN. All relevant data are available in the main text, in the Supporting Information, or from the authors.

**Acknowledgements**

This work received support from National Key Research and Development Program of China (2021YFA1401000), National Natural Science Foundation of China (62435009), and Beijing Natural Science Foundation (Z220008).

**Author contributions**

J.H., Y.D. and Y.L. conceived the idea. J.H., J.Q. and Y.D. performed the experiments. J.H. and Y.D. performed the numerical simulations and analysed the data. J.W., M.L. and X.C. designed and fabricated the integrated photonics processing unit. J.W., M.L., J.H., G.Y., J.Q. and X.C. characterized the device. J.H. and Y.L. wrote the manuscript with contributions from J.Q., Y.D., T.D., R.Z., S.Y., Z.L., X.C., Y.Y. and G.W. X.C., Y.Y., G.W. and Y.L. supervised the research.

**Competing interests**

The authors declare no competing interests.

**Data and materials availability**

The data that supports the findings of this study are available from the corresponding authors upon reasonable request.

**Supplementary information**

Supplementary Text, Supplementary Figures 1-11, and Supplementary Tables 1-4